\documentclass[onecolumn,showpacs,preprintnumbers,amsmath]{revtex4}
\usepackage{graphicx}
\usepackage{dcolumn}
\usepackage{bm}
\usepackage{epsfig}
\usepackage{subfigure}

\newcommand{\bi}{\begin{itemize}}
\newcommand{\ei}{\end{itemize}}
\newcommand{\be}{\begin{eqnarray}}
\newcommand{\ee}{\end{eqnarray}}
\newcommand{\bbmatrix}{\left( \begin{array}}
\newcommand{\eematrix}{\end{array} \right)}

\begin{document}
\title{U(1) $\times$ U(1) / Z$_2$ Kosterlitz-Thouless transition of the Larkin-Ovchinnikov phase in an anisotropic two-dimensional system}
\author{Chungwei Lin, Xiaopeng Li, and W.~Vincent Liu}
\affiliation{Department of Physics and Astronomy, University of
  Pittsburgh, Pittsburgh, PA 15260}
\affiliation{Kavli Institute for Theoretical Physics, University of
  California, Santa Barbara, CA 93106} 

\begin{abstract}
We study Kosterlitz-Thouless (KT) transitions of the
Larkin-Ovchinnikov (LO) phase for a two-dimensional system composed of
coupled one-dimensional tubes of fermions.  
  The low energy excitations involve the oscillation of the LO
stripe and the fluctuation of the phase, which can be
described by an effective theory composed of two anisotropic XY models.  
We compute from a microscopic
model the coefficients of the XY models from which the KT transition
temperatures are determined.  
We found the $T^{KT} \propto t_{\perp}$
for small intertube tunneling $t_{\perp}$  and over a wide range of parameters
the fractional defects costs least energy to proliferate. 
As $t_{\perp}$ increases
the system undergoes a first-order transition to the normal phase at zero
temperature.
Our method  is efficient and
can be used to determine the Goldstone excitations of any stripe order involving charge or spin degrees of freedom.
\end{abstract}
\preprint{NSF-KITP-10-149}
\pacs{03.75.Hh, 03.75.Ss, 67.85-d }
\maketitle


The role of topological excitations of striped superconducting states
has been intensively studied
\cite{KT_1973,PhysRevB.79.064515,Agterberg_08_NatPhys,PhysRevLett.103.010404,Berg_09_NatPhys}
since at finite temperature the proliferation of those defects can lead to
possible exotic phases, such as the charge $4$ superfluid
\cite{Berg_09_NatPhys,PhysRevLett.95.266404}.
A typical striped superconducting state is the
Fulde-Ferrell-Larkin-Ovchinnikov (FFLO)
\cite{PhysRev.135.A550,larkin:1964zz} state which is believed to exist
in heavy-fermion superconductor CeCoIn$_5$
\cite{Radovan_03,PhysRevLett.91.187004} and  has been recently
proposed to occur in the system involving  $p$-orbital bands \cite{PhysRevA.82.033610,Cai_10}.  
Since the FFLO
order is more likely to occur in the quasi-one-dimensional (1D) system
\cite{PhysRevLett.99.250403}, the cold atom system with two imbalanced
species of atoms confined in a lattice array of 1D tubes formed by
coherent laser beams \cite{Liao_10} seems more promising to display
the direct evidence. Since the intertube coupling can be tuned
relatively with ease in the cold atom system by controlling the
intensity of trapping lasers, it is suitable to study the dimensional
crossover phenomena \cite{Carlson_00,Biermann_01,Ho_04,Kollath_08}.

\begin{figure}
   \centering
   \epsfig{file=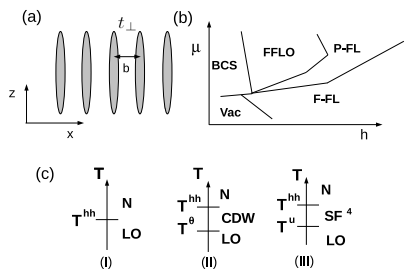, width=0.3 \textwidth}
   \caption{(a) Configuration of the system: arrays of 1D tubes with intertube distance $b$ and intertube tunneling $t_{\perp}$. 
      (b) A schematic plot of quasi-1D phase diagram as a function of $\mu$ and $h$. Vac: vacuum state (no particle);
      P-FL: partially polarized Fermi liquid; F-FL: fully polarized Fermi liquid. Our study here focuses on the FFLO regime. 
      (c) Possible phases as a function of temperature. N:normal Fermi liquid; CDW: charge density wave;
    SF$^{4}$: charge 4 superfluid.}
   \label{fig:config_phases}
\end{figure}

\begin{figure}
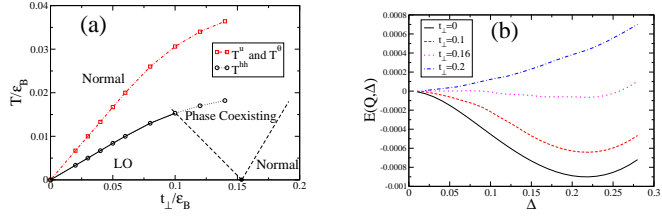

   \centering
   \subfigure{\epsfig{file = Tc_mu2_h1.115_tp0-0.14_Q1.2_Gap0.22.eps,  width = 0.22\textwidth}}
    \qquad
   \subfigure{\epsfig{file = E_gap_U1.000_mu2.00_h1.15_Gap0.0-0.3_Q1.20_mz0.00_tp0-0.2_dkz0.05.eps,  width = 0.22\textwidth}}
  \caption{ (Color online) (a) Phase diagram for $\mu=2$, $h=1.145$, $Q=1.2$, $\Delta_0 = 0.22$. Curves with ligands are computed whereas
  two dashed lines embracing the phase coexisting region are schematic. The calculated transition temperature $T^{hh}$
   within the phase coexisting region  is not well defined and its plot is switched to the dotted line with circles. 
  (b) Energy as a function of gap amplitude $\Delta$ for $\mu=2$, $h=1.145$, $Q=1.2$, $t_{\perp}=0-0.2$.
    The energy minimum occurs at $\Delta = 0.22$ for $t_{\perp}<0.15$. }
   \label{fig:Tc_a0}
\end{figure}

Numerous exotic phases have been predicted from effective field theories \cite{Agterberg_08_NatPhys}, 
but the phase diagram of these exotic phases is not established for cold atom experiments yet. 
In cold atom experiments, the microscopic parameters (like interaction strength) are tunable and measurable, and this motivates our detailed 
study of the Kosterlitz-Thouless (KT) transitions of the Larkin-Ovchinnikov (LO) phase starting from a microscopic model. 
In this work we study a quasi-1D two-dimensional (2D) spinful fermionic system composed of coupled 1D tubes as illustrated in Fig. \ref{fig:config_phases}a
where at zero temperature the LO order is the ground state.
We determine the Kosterlitz-Thouless temperature of LO phase
(the FFLO regime in Fig. \ref{fig:config_phases}b \cite{PhysRevLett.99.250403}) as a function of intertube coupling $t_{\perp}$ from
a microscopic model. We found that  over a wide range of parameters
the fractional defect dominates the phase transition and transition temperature
 is linear in $t_{\perp}$ for small $t_{\perp}$ (Fig. \ref{fig:Tc_a0}a). 
At zero temperature the transition from LO to normal phase,  driven by the disappearance 
of the Fermi surface nesting upon increasing $t_{\perp}$, is of first order
(Fig. \ref{fig:Tc_a0}b).

Before introducing the microscopic model, we first discuss the Goldstone modes of LO phase 
\cite{Agterberg_08_NatPhys,PhysRevLett.103.010404,Berg_09_NatPhys} from the symmetry point of view.
The LO phase is characterized by an  order parameter of  stripe
  configuration 
\be
\Delta(x,z) \propto \langle  
c_{\downarrow}(x,z) c_{\uparrow}(x,z) \rangle \propto \Delta_0 f(z)
\ee
where $\Delta_0$  is the amplitude of the order parameter while  $f(z)$
describes  the stripe satisfying $f(z+L/2)=-f(z)$ and $f(z)=f(-z)$ in our coordinate choice. 
The LO wave vector $Q$ is defined as $Q\equiv 2 \pi/L$ with $L$ the period of the stripe. 
Since the LO phase breaks both translational and $U(1)$ symmetries, it has
two branches of Goldstone modes -- the oscillation of the stripe,
and the phase fluctuation of the amplitude. Under these fluctuations, the order parameter becomes 
\be
\Delta(x,z) =\Delta_0 e^{i \theta(x,z)} \,f( z+ u(x,z) )  
\label{eqn:f(z)}
\ee
where $u(x,z)$, $\theta(x,z)$ are generalized elastic fields \cite{Chaikin} to describe the Goldstone modes. 
Physically $u$ represents the  small oscillation  of the stripe LO order whereas
$\theta$ the phase fluctuation of the 
amplitude. In the quasi-1D system, $x$ and $z$ directions are not equivalent. Therefore to the quadratic order
the  total free energy in terms of generalized elastic fields are described by 
two anisotropic XY models \cite{PhysRevLett.89.095701,Berg_09_NatPhys}
\be
\Delta F = \int dx dz  \left[\frac{A}{2} (Q  u_x)^2 + \frac{B}{2} (Q  u_z)^2 
+ \frac{C}{2} \theta_x ^2 + \frac{D}{2} \theta_z^2 \right]
\label{eqn:Free}
\ee
where $f_x = \partial_x f $. In our notation, $u$ and $\theta/Q$ have the dimension of length, 
their  first derivatives are dimensionless, and  
coefficients $A$, $B$, $C$, $D$ have the dimension of energy.
For results presented in this paper, we take $f(z) = \cos Qz$ which is $(e^{i Qz} + e^{-i Qz})/2$. 
In this case, these two Goldstone modes
correspond to phase fluctuations of two Fulde-Ferrell (FF) order $\Delta(x,z)= \Delta_0 (e^{i Q(z+u^+)} + e^{-i Q(z+u^-)})/2$. For 
the FF order, $\Delta(x,z)\propto e^{i Qz}$ which does not break
translation symmetry.  When identifying
$u^+=u+\theta/Q$, $u^-=u-\theta/Q$, $\Delta(x,z)=\Delta_0 \cos( z+u(x,z) ) \,\, e^{i \theta(x,z)}$, consistent with the functional form
in Eq. (\ref{eqn:f(z)}). To be general, we shall not specify the form of $f(z)$ unless necessary.

In 2D, each elastic field is associated with one topological defect.
For $u$ the defect is the (edge) dislocation satisfying $\oint \vec{\nabla} u \cdot d \vec{l} = L n_d$;
for $\theta$ the defect is the vortex satisfying $\oint \vec{\nabla} \theta \cdot d \vec{l} = 2 \pi n_v$ with $n_d,n_v$ integers.
There is another topological defect referred to as a half-vortex half-dislocation (HH) 
where $(n_d,n_v) = (\pm 1/2, \pm 1/2)$, which originates from
the $Z_2$ symmetry of the order parameter \cite{Berg_09_NatPhys,PhysRevLett.103.010404} -- when circulating around an HH defect,
each of the half vortex and the half dislocation introduces a minus sign leaving the order parameter unchanged.
The proliferation of topological defects leads to Kosterlitz-Thouless (KT) transition.
The KT transition temperature $T^u = \frac{\pi}{2} \sqrt{AB}$ for dislocations,
$T^{\theta} =\frac{\pi}{2} \sqrt{CD}$ for vortices, and $T^{hh}=\frac{\pi}{8} (\sqrt{AB} +\sqrt{CD}) = (T^u+T^{\theta})/4$ for HH 
\cite{Berg_09_NatPhys}. The last temperature cannot be highest. When increasing the temperature, there are three distinct possibilities
as illustrated in Fig. \ref{fig:config_phases}c: (I) $T^{hh}$ is the lowest, (II) $T^{\theta}$ the lowest,  and (III) $T^{u}$
the lowest \cite{Berg_09_NatPhys}. 
For (I) there is only one transition from LO to normal state at $T^{hh}$. 
For (II), the LO phase first becomes a charge density wave (CDW) state at $T^{\theta}$ and then normal at $T^{hh}$.
For (III),the LO phase first becomes a charge $4$ superfluid at $T^{u}$ and then normal at $T^{hh}$.

The microscopic model we use is a one band model with attractive contact interaction in a quasi-1D system \cite{PhysRevLett.99.250403}.
\be
H &=& \sum_{\vec{k},\sigma} \xi_{\vec{k},\sigma} c^{\dagger}_{\vec{k},\sigma} c_{\vec{k},\sigma} \nonumber \\ 
&+& g_{1D} \sum_{i_x} \int dz\, c^{\dagger}_{i_x, \uparrow}(z) c^{\dagger}_{i_x, \downarrow}(z) c_{i_x, \downarrow}(z) c_{i_x, \uparrow}(z)
\ee
with $\xi_{\vec{k},\sigma} = \frac{\hbar^2 k_z^2}{2m} - 2t_{\perp} \cos (b k_x) - \mu + h (-1)^{\sigma}$,
where $k_z$ is unbound, $|b k_x|<\pi$ with $b$ the intertube distance, and $(-1)^{\uparrow}=1$, $(-1)^{\downarrow}=-1$.
Following Ref \cite{PhysRevLett.98.070402,PhysRevLett.99.250403}, we measure all lengths in the 1D scattering
length $a_{1D} = -\frac{2 \hbar^2}{m g_{1D}}$ and all energies by the 1D bound energy $\epsilon_B=\frac{\hbar^2}{m} \frac{1}{a^2_{1D}}$.
The dimensionless parameters in this model are $t_{\perp}/\epsilon_B$, $\mu/\epsilon_B$,  $h/\epsilon_B$, and $b/a_{1D}$.
The attractive interaction implies negative $g_{1D}$ and thus positive $a_{1D}$. 
The relation between $a_{1D}$ and $a_{3D}$ is \cite{PhysRevLett.81.938}
$a_{1D} = -a_{\perp} \left( \frac{a_{\perp}}{a_{3D}} - \frac{1.4603}{\sqrt{2}} \right)$
with $a_{\perp}=\sqrt{ \frac{\hbar}{m \omega_{\perp}} }$.
The typical $a_{\perp}$ is of the order 100nm and $a_{3D}$ can be controlled by the Fashbach resonance.
Taking $a_{1D} = 100$nm, $m=6/(6\times 10^{23})$g ($^6$Li) \cite{Liao_10}, the bound state energy is
$\epsilon_B \sim 1.5 \times 10^{-6}$K. 

To obtain the effective theory at given parameters,
we first solve the Hamiltonian by the variational method where the order parameter is assumed to be sinusoidal 
$\Delta_{i_x}(z) = \Delta(z) = \Delta_0 \cos(Q z) = g_{1D} \langle c_{i_x, \downarrow}(z) c_{i_x, \uparrow}(z) \rangle $.
$f(z)$ in Eq. (\ref{eqn:f(z)}) is chosen to be $\cos Q z$.
The mean field  (Bogoliubov de Gennes) Hamiltonian is
\be
H_{mf} &=&
\sum_{\vec{k}} (c^{\dagger}_{\vec{k}, \uparrow}, c_{-\vec{k}, \downarrow}) H(\Delta_0.Q) 
\bbmatrix{c} c_{\vec{k}, \uparrow} \\ c^{\dagger}_{-\vec{k}, \downarrow} \eematrix  \nonumber \\
&+& \sum_{\vec{k}} \xi_{\vec{k},\downarrow}  + L_z N_x\frac{\Delta_0^2}{2 |g_{1D}|}
\ee
where
\be
H(\Delta_0,Q) = \bbmatrix{cc} \xi_{\vec{k},\uparrow} & B  \\
B & -\xi_{-\vec{k}, \downarrow}
\eematrix
\ee
with the block $B = \frac{\Delta_0}{2} \delta_{k_x,-k_x} (\delta_{k_z+Q, -k_z} +  \delta_{k_z-Q, -k_z})$.
The $\Delta_0$ and $Q$ are determined by minimizing the free energy with respect to $Q$ and $\Delta_0$ given by
\be
F[\Delta] = -T \sum_n \log[1 + e^{\frac{-\epsilon_n}{T} }]  
+ \sum_{\vec{k}} \xi_{\vec{k},\downarrow} +  L_z N_x \frac{\Delta_0^2}{2 |g_{1D}|}
\label{eqn:e_Del}
\ee 
where $\epsilon_n$ are $all$ eigenvalues of the matrix $H(\Delta_0, Q)$.
In the $T\rightarrow 0$ limit, $-T \sum_n \log[1 + e^{-\epsilon_n/T}] =\sum_n \epsilon_n \Theta(-\epsilon_n)  $.

The energy cost for given strain configurations $u(x,z)$, $\theta(x,z)$ of elastic fields   
is computed by $\Delta F(b) = F[\Delta_0 e^{i  \theta} f (z+u)] - F[\Delta_0 f(z)]$, which to the lowest order 
of $u$ and $\theta$ in quasi-1D reduces to the form
\be
\Delta F(b) &=& b \sum_{x_i} \int dz 
\left[ \frac{A(b)}{2} (Q u_x)^2 + \frac{B(b)}{2} (Q u_z)^2 \right. \nonumber \\
&& \left. + \frac{C(b)}{2} \theta_x^2 + \frac{D(b)}{2} \theta_z^2\right] 
\label{eqn:F_u}
\ee
where in the quasi-1D system $u_x (x_i) \equiv \frac{u(x_i+b) - u(x_i)}{b}$, $\theta_x (x_i) \equiv \frac{\theta(x_i+b) - \theta(x_i)}{b}$.
Replacing  $b\sum_{i_x} \rightarrow \int dx$, Eq. (\ref{eqn:F_u}) is the same as Eq. (\ref{eqn:Free}). 

We now show that the KT transition temperature is independent of the intertube distance $b$ when $t_{\perp}$ is fixed. 
For simplicity we only consider the stripe oscillation field $u$. The same argument applies to the phase field $\theta$.  
When $b \rightarrow \alpha b$, the total free energy cost due to compression or stretching along $z$ is unchanged
(since the energy depends only on $t_{\perp}$ which is fixed)   but the energy density changes. Consequently
$\Delta F_z(b) = \Delta F_z(\alpha b)$ implies
\be
 b N_x L_z \frac{B(b)}{2} (Q u_z)^2 = \alpha b N_x L_z \frac{B(\alpha b)}{2} (Q u_z)^2
\ee
leading to $B(\alpha b) = B(b)/\alpha$. The free energy caused by $u_x$ depends only on intertube coupling $t_{\perp}$ and therefore
the elastic field difference  between two adjacent tubes $u(x+b)-u(x)$. 
When $b \rightarrow \alpha b$, as long as $u(x+b)-u(x) = u(x+\alpha b)-u(x)$ (same $t_{\perp}$) the total free 
energy remains unchanged which leads to 
\be
&&\frac{A(b)}{2} L_z b \sum_x Q^2 \left(\frac{u(x+b)-u(x)}{b} \right)^2 \nonumber \\ &=&
\frac{A(\alpha b)}{2} L_z (\alpha b) \sum_x Q^2 \left(\frac{u(x+\alpha b)-u(x)}{\alpha b} \right)^2
\ee
which leads to $A(\alpha b) =\alpha A(b)$. Their product $A(b) B(b)$ is independent of $b$,
so is the  transition temperature $T^{KT} \propto \sqrt{A B}$.
Since our main interest is the transition temperature, we take $b=a_{1D}=1$.

To obtain coefficients $A$, $B$, $C$, $D$ in Eq. (\ref{eqn:Free}), we take the following approach.
Take $B$ as an example, we choose $u(x,z) = u_z z$, $\theta(x,z)=0$, compute $\Delta F(b)$ for several $u_z$,
and fit $\delta F(b,u_z) \equiv \Delta F(b; u_z)/(b N_x L_z) =  \frac{B(b)}{2} (Q u_z)^2$ as expressed in Eq. (\ref{eqn:F_u}). 
The same procedure apply to $A$, $C$, $D$. There is another approach to obtain these coefficients involving
Green's function \cite{PhysRevB.81.224507} which requires computing the inverse of a matrix and is very time-consuming. 
Our approach instead only involves the computation of eigenvalues \cite{PhysRev.187.556} which allows us to
include more $k$-points. 

\begin{figure}[htbp]
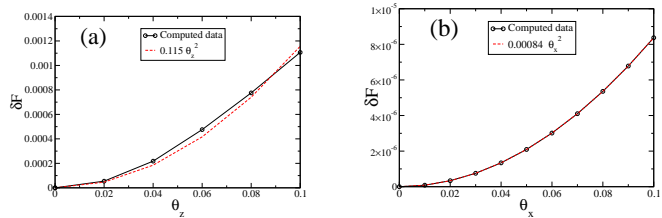

\begin{center}
   \subfigure{\epsfig{file = E_vz_Gap0.220_mu2.00_h1.15_vz0.0-0.2_Q1.20_mz0.00_tp0.10_dkz0.05_kmax18.1.eps,  width = 0.22\textwidth}}
\qquad
   \subfigure{\epsfig{file = E_vx_Gap0.220_mu2.00_h1.15_vx0.0-0.1_Q1.20_mz0.00_tp0.10_dkz0.05_kmax24.1.eps,  width = 0.22\textwidth}}
   \caption{(Color online) Energy density as a function of $\theta_z$ and $\theta_x$ for $\mu=2$, $h=1.145$, $Q=1.2$, $\Delta_0 = 0.22$..  }
   \label{fig:F_theta}
\end{center}
\end{figure}

Some key steps of computing $\delta F(b,u_z)$ are are summarized here. 
To obtain the coefficients $C$ and $D$ requires computing the eigenvalues of 
\be
H(\theta)=\bbmatrix{cc} (h_0-\mu)+h & \Delta_0 f(z) e^{i \theta}  \\
\Delta_0 f(z) e^{-i  \theta} & -(h_0-\mu)+h
\eematrix
\ee
where $h_0$ is diagonalized in momentum space as $\frac{\hbar^2 k_z^2}{2m} - 2t_{\perp} \cos b k_x$.
By performing a local gauge transformation $c(z)\rightarrow c(z)e^{-i \theta/2}$, $H(\theta)$
in new coordinate becomes $H_1(\theta)$  \cite{Nagaosa,PhysRevB.81.224507} which is
\be
H_1(\theta) = \bbmatrix{cc} \bar{h}_{0,\uparrow} & \Delta_0 f(z) \\
\Delta_0 f(z)  & -\bar{h}_{0,\downarrow} \eematrix
\ee
where $\bar{h}_{0,\uparrow}$ and $\bar{h}_{0,\downarrow}$ are diagonalized in momentum space as $\xi^+(\vec{k}) = \xi_{\vec{k}+\vec{v}/2,\uparrow}$ and $\xi^-(-\vec{k}) = \xi_{-\vec{k}+\vec{v}/2,\downarrow}$ respectively with $ \vec{v} = \vec{\nabla}\theta/2$. 
The eigenvalues of $H$ and $H_1$ are identical, but this transformation automatically
obtains the derivative of $\theta$, i.e. $\theta_x$ and $\theta_z$, in the diagonal blocks.
We use $\mu=2$, $h=1.145$, $t_{\perp}=0.1$ as an example. Minimizing the energy functional
with respect to $\Delta_0$ and $Q$ leads to $Q=1.2$, $\Delta_0 = 0.22$. 
Fig. \ref{fig:F_theta} shows
$\delta F (\theta_z)$ for $\theta(x,z) = \theta_z z$ and $\delta F (\theta_x)$ for $\theta(x,z) = \theta_x x$
from which the quadratic fit leads to $C=0.00168$ and $D=0.23$. 

\begin{figure}[htbp]
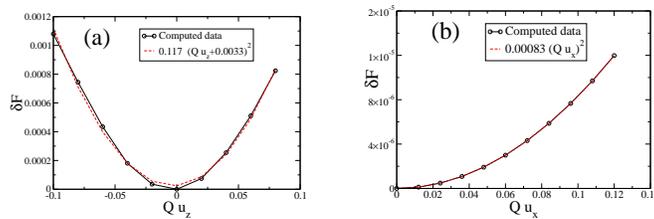

\begin{center}
   \subfigure{\epsfig{file = E_uz_mu2.00_h1.15_Gap0.22_Q1.20_tp0.10_dkz0.05_uz-0.10-0.10.eps,  width = 0.22\textwidth}}
\qquad
   \subfigure{\epsfig{file = F_u_mu2.00_h1.15_Gap0.22_Q1.20_mz0.00_tp0.10_dkz0.05_u0.00-0.10.eps,  width = 0.22\textwidth}}
   \caption{(Color online) Energy density as a function of $u_z$ and $u_x$ for  $\mu=2$, $h=1.145$, $Q=1.2$, $\Delta_0 = 0.22$.  }
   \label{fig:F_u}
\end{center}
\end{figure}

To obtain the coefficients $A$ and $B$ requires computing the eigenvalues of
\be
H(u)=\bbmatrix{cc} (h_0-\mu)+h & \Delta_0 f(z+ u)   \\
\Delta_0 f(z+ u) & -(h_0-\mu)+h
\eematrix.
\ee
A useful trick is to do the calculation in a new coordinate whose order parameter is exactly $\Delta_0 f(z)$\cite{PhysRevB.81.224507}.
We stress here the Jacobian arising from the coordinate transform has to be taken into account
because it is the free energy, not the free energy density,  which is invariant under 
coordinate transformation. 
Again we use parameter $\mu=2$, $h=1.145$, $t_{\perp}=0.1$ as an example.
Fig. \ref{fig:F_u} shows $\delta F (Q u_z)$ for $u(x,z) = u_z z$ and $\delta F (Q u_x)$ for $\theta(x,z) = u_x x$
from which we can fit $A=0.00168$ and $B=0.234$. 
Note that $f(z)$ is taken to be $\cos Qz$ for the results presented here. However we emphasize that the coefficients $A, B, C, D$
can be obtained for any given order parameters.

We compute the coefficients for several $t_{\perp}$ and determine all three KT transition temperatures.  
Our main result is shown in Fig. \ref{fig:Tc_a0}a where the phase diagram as a function of $T$ and $t_{\perp}$ is plotted 
for a representative set of parameters $\mu=2$, $h=1.145$, $Q=1.2$, $\Delta_0 = 0.22$.
When $t_{\perp} = 0$, there is no intertube coupling and no correlation along $x$ leading to zero $T^{KT}$.
As the system goes from pure 1D to quasi-1D, $T^{KT}(t_{\perp}) \propto t_{\perp}$.
More specifically, we found the coefficients associated with $z$ derivative, i.e. $B$ and $D$, depend very weakly on $t_{\perp}$
whereas those with $x$ derivative, i.e. $A$ and $C$, depend quadratically on $t_{\perp}$. This explains the linear $t_{\perp}$ dependence of
$T^{KT}$ ($\propto \sqrt{AB}, \sqrt{CD}$).  We found $T^{\theta}$ and $T^u$ are very close because
the ``cosine'' ansatz is very close to two $decoupled$ FF order with opposite wave vector for small $\Delta_0$ and
it is the coupling between $u^+$ and $u^-$ (the fluctuations of two FF orders) which lifts the degeneracy of $u$ and $\theta$ fields.
In this case the proliferation of half-vortex half-dislocation 
costs the least energy and the system only undergoes one transition from LO to normal at $T^{hh}$ when raising the temperature, 
as shown in the case I of Fig. \ref{fig:config_phases}c.  
The transition temperature is of the order of 0.03 $\epsilon_B$ which is  roughly $8\times10^{-8}$K for the system of $^6$Li with $a_{1D}=100$nm.
One also notes that the obtained $T^{KT} (\sim 0.03) $ is an order of magnitude smaller than the mean field gap $\Delta_0 (=0.22)$,
so the coefficients computed at $T=0$ are almost identical (less than $1\%$ difference) to those computed at $T \sim T^{KT}$.

At $T=0$, our simulation suggests the quantum phase transition from LO to normal phases upon increasing $t_{\perp}$ is of first order.
Fig. \ref{fig:Tc_a0}b shows the energy as a function of $\Delta$ for $\mu=2$, $h=1.145$, $Q=1.2$, $t_{\perp}\in (0,0.2)$
where the minimum determines the value of $\Delta_0$. We found that as $t_{\perp}$ increases $\Delta_0$
stays around 0.22 and when $t_{\perp}>0.16$ $\Delta_0$ becomes zero. Around $t_{\perp}=0.16$,
the $E(\Delta)$ is essentially flat  with several shallow minima. 
We note that the FFLO to normal transition as a function of temperature (fixed $\mu$, and $h$) \cite{Radovan_03,PhysRevLett.91.187004}  
or $h$ (fixed $\mu$, $T$)  \cite{Matsou_98} is also of first order.
At finite temperature around the critical $t_{\perp}$, the system 
is in the phase coexisting region. 


We have assumed a sinusoidal order parameter  in the current calculation.
However near BCS/LO transition (Fig. \ref{fig:config_phases}b) \cite{PhysRevLett.98.070402,PhysRevLett.99.250403}, 
the order parameter behaves more domain-wall like \cite{Matsou_98,PhysRevA.75.063601} than sinusoidal. Therefore
the ansatz with sinusoidal order parameter does not capture all physics. Close to the BCS/LO transition, we expect that
the stripe fluctuation should be stronger than the phase fluctuation ($A<C$, $B<D$) and a two-stage transition
with charge 4 superfluid shown as the case III in Fig. \ref{fig:config_phases}c can happen. In a cold atom trap
where the chemical potential is a position-dependent, the interface between  phases shown in Fig. \ref{fig:config_phases}c
is unavoidable and worth investigating.

In summary, we have computed from a microscopic model the effective theories of Goldstone modes  of the LO order 
for a quasi-1D fermionic system from which the Kosterlitz-Thouless transition temperatures are determined. 
The transition temperatures are found to depend linearly on the intertube coupling $t_{\perp}$.   However,
the method applied here neglects the quantum fluctuation along the 1D tubes which can modify this linear $t_{\perp}$
dependence, especially at $t_{\perp} \rightarrow 0$ \cite{PhysRevA.78.063605}. 
As $t_{\perp}$ increases,
the system goes to a phase coexisting regime sandwiched by the LO and normal phases. 
Our approach can generally determine the Goldstone excitations of any stripe order involving charge or spin
from a microscopic model which should be useful for comparison between theories and experiments.

We thank Erhai Zhao, Zixu Zhang, Han Pu, and David Huse for very helpful
discussions.  This work is supported under ARO Award
No. W911NF-07-1-0464 with funds from the DARPA OLE Program and ARO
Award No. W911NF-07-1-0293. We thank the KITP of UCSB where this
research is supported in part by the NSF under
Grant No. PHY05-51164.

\bibliography{Q1D_LO_melting_8}

\begin{thebibliography}{28}
\expandafter\ifx\csname natexlab\endcsname\relax\def\natexlab#1{#1}\fi
\expandafter\ifx\csname bibnamefont\endcsname\relax
  \def\bibnamefont#1{#1}\fi
\expandafter\ifx\csname bibfnamefont\endcsname\relax
  \def\bibfnamefont#1{#1}\fi
\expandafter\ifx\csname citenamefont\endcsname\relax
  \def\citenamefont#1{#1}\fi
\expandafter\ifx\csname url\endcsname\relax
  \def\url#1{\texttt{#1}}\fi
\expandafter\ifx\csname urlprefix\endcsname\relax\def\urlprefix{URL }\fi
\providecommand{\bibinfo}[2]{#2}
\providecommand{\eprint}[2][]{\url{#2}}

\bibitem[{\citenamefont{Kosterlitz and Thouless}(1973)}]{KT_1973}
\bibinfo{author}{\bibfnamefont{J.~M.} \bibnamefont{Kosterlitz}}
  \bibnamefont{and} \bibinfo{author}{\bibfnamefont{D.}~\bibnamefont{Thouless}},
  \bibinfo{journal}{J. Phys. C} \textbf{\bibinfo{volume}{6}},
  \bibinfo{pages}{1181} (\bibinfo{year}{1973}).

\bibitem[{\citenamefont{Berg et~al.}(2009{\natexlab{a}})\citenamefont{Berg,
  Fradkin, and Kivelson}}]{PhysRevB.79.064515}
\bibinfo{author}{\bibfnamefont{E.}~\bibnamefont{Berg}},
  \bibinfo{author}{\bibfnamefont{E.}~\bibnamefont{Fradkin}}, \bibnamefont{and}
  \bibinfo{author}{\bibfnamefont{S.~A.} \bibnamefont{Kivelson}},
  \bibinfo{journal}{Phys. Rev. B} \textbf{\bibinfo{volume}{79}},
  \bibinfo{pages}{064515} (\bibinfo{year}{2009}{\natexlab{a}}).

\bibitem[{\citenamefont{Agterberg and Tsunetsugu}(2008)}]{Agterberg_08_NatPhys}
\bibinfo{author}{\bibfnamefont{D.~F.} \bibnamefont{Agterberg}}
  \bibnamefont{and}
  \bibinfo{author}{\bibfnamefont{H.}~\bibnamefont{Tsunetsugu}},
  \bibinfo{journal}{Nat. Phys.} \textbf{\bibinfo{volume}{4}},
  \bibinfo{pages}{639} (\bibinfo{year}{2008}).

\bibitem[{\citenamefont{Radzihovsky and
  Vishwanath}(2009)}]{PhysRevLett.103.010404}
\bibinfo{author}{\bibfnamefont{L.}~\bibnamefont{Radzihovsky}} \bibnamefont{and}
  \bibinfo{author}{\bibfnamefont{A.}~\bibnamefont{Vishwanath}},
  \bibinfo{journal}{Phys. Rev. Lett.} \textbf{\bibinfo{volume}{103}},
  \bibinfo{pages}{010404} (\bibinfo{year}{2009}).

\bibitem[{\citenamefont{Berg et~al.}(2009{\natexlab{b}})\citenamefont{Berg,
  Fradkin, and Kivelson}}]{Berg_09_NatPhys}
\bibinfo{author}{\bibfnamefont{E.}~\bibnamefont{Berg}},
  \bibinfo{author}{\bibfnamefont{E.}~\bibnamefont{Fradkin}}, \bibnamefont{and}
  \bibinfo{author}{\bibfnamefont{S.~A.} \bibnamefont{Kivelson}},
  \bibinfo{journal}{Nat. Phys.} \textbf{\bibinfo{volume}{5}},
  \bibinfo{pages}{830} (\bibinfo{year}{2009}{\natexlab{b}}).

\bibitem[{\citenamefont{Wu}(2005)}]{PhysRevLett.95.266404}
\bibinfo{author}{\bibfnamefont{C.}~\bibnamefont{Wu}}, \bibinfo{journal}{Phys.
  Rev. Lett.} \textbf{\bibinfo{volume}{95}}, \bibinfo{pages}{266404}
  (\bibinfo{year}{2005}).

\bibitem[{\citenamefont{Fulde and Ferrell}(1964)}]{PhysRev.135.A550}
\bibinfo{author}{\bibfnamefont{P.}~\bibnamefont{Fulde}} \bibnamefont{and}
  \bibinfo{author}{\bibfnamefont{R.~A.} \bibnamefont{Ferrell}},
  \bibinfo{journal}{Phys. Rev.} \textbf{\bibinfo{volume}{135}},
  \bibinfo{pages}{A550} (\bibinfo{year}{1964}).

\bibitem[{\citenamefont{Larkin and Ovchinnikov}(1964)}]{larkin:1964zz}
\bibinfo{author}{\bibfnamefont{A.~I.} \bibnamefont{Larkin}} \bibnamefont{and}
  \bibinfo{author}{\bibfnamefont{Y.~N.} \bibnamefont{Ovchinnikov}},
  \bibinfo{journal}{Zh. Eksp. Teor. Fiz.} \textbf{\bibinfo{volume}{47}},
  \bibinfo{pages}{1136} (\bibinfo{year}{1964}).

\bibitem[{\citenamefont{Radovan et~al.}(2003)\citenamefont{Radovan, Fortune,
  Murphy, Hannahs, Palm, Tozer, and Hall}}]{Radovan_03}
\bibinfo{author}{\bibfnamefont{H.~A.} \bibnamefont{Radovan}},
  \bibinfo{author}{\bibfnamefont{N.~A.} \bibnamefont{Fortune}},
  \bibinfo{author}{\bibfnamefont{T.~P.} \bibnamefont{Murphy}},
  \bibinfo{author}{\bibfnamefont{S.~T.} \bibnamefont{Hannahs}},
  \bibinfo{author}{\bibfnamefont{E.~C.} \bibnamefont{Palm}},
  \bibinfo{author}{\bibfnamefont{S.~W.} \bibnamefont{Tozer}}, \bibnamefont{and}
  \bibinfo{author}{\bibfnamefont{D.}~\bibnamefont{Hall}},
  \bibinfo{journal}{Nature} \textbf{\bibinfo{volume}{425}}, \bibinfo{pages}{51}
  (\bibinfo{year}{2003}).

\bibitem[{\citenamefont{Bianchi et~al.}(2003)\citenamefont{Bianchi, Movshovich,
  Capan, Pagliuso, and Sarrao}}]{PhysRevLett.91.187004}
\bibinfo{author}{\bibfnamefont{A.}~\bibnamefont{Bianchi}},
  \bibinfo{author}{\bibfnamefont{R.}~\bibnamefont{Movshovich}},
  \bibinfo{author}{\bibfnamefont{C.}~\bibnamefont{Capan}},
  \bibinfo{author}{\bibfnamefont{P.~G.} \bibnamefont{Pagliuso}},
  \bibnamefont{and} \bibinfo{author}{\bibfnamefont{J.~L.}
  \bibnamefont{Sarrao}}, \bibinfo{journal}{Phys. Rev. Lett.}
  \textbf{\bibinfo{volume}{91}}, \bibinfo{pages}{187004}
  (\bibinfo{year}{2003}).

\bibitem[{\citenamefont{Zhang et~al.}(2010)\citenamefont{Zhang, Hung, Ho, Zhao,
  and Liu}}]{PhysRevA.82.033610}
\bibinfo{author}{\bibfnamefont{Z.}~\bibnamefont{Zhang}},
  \bibinfo{author}{\bibfnamefont{H.-H.} \bibnamefont{Hung}},
  \bibinfo{author}{\bibfnamefont{C.~M.} \bibnamefont{Ho}},
  \bibinfo{author}{\bibfnamefont{E.}~\bibnamefont{Zhao}}, \bibnamefont{and}
  \bibinfo{author}{\bibfnamefont{W.~V.} \bibnamefont{Liu}},
  \bibinfo{journal}{Phys. Rev. A} \textbf{\bibinfo{volume}{82}},
  \bibinfo{pages}{033610} (\bibinfo{year}{2010}).

\bibitem[{\citenamefont{Cai et~al.}(2010)\citenamefont{Cai, Wang, and
  Wu}}]{Cai_10}
\bibinfo{author}{\bibfnamefont{Z.}~\bibnamefont{Cai}},
  \bibinfo{author}{\bibfnamefont{Y.}~\bibnamefont{Wang}}, \bibnamefont{and}
  \bibinfo{author}{\bibfnamefont{C.}~\bibnamefont{Wu}} (\bibinfo{year}{2010}),
  \eprint{arXiv:1009.3257}.

\bibitem[{\citenamefont{Parish et~al.}(2007)\citenamefont{Parish, Baur,
  Mueller, and Huse}}]{PhysRevLett.99.250403}
\bibinfo{author}{\bibfnamefont{M.~M.} \bibnamefont{Parish}},
  \bibinfo{author}{\bibfnamefont{S.~K.} \bibnamefont{Baur}},
  \bibinfo{author}{\bibfnamefont{E.~J.} \bibnamefont{Mueller}},
  \bibnamefont{and} \bibinfo{author}{\bibfnamefont{D.~A.} \bibnamefont{Huse}},
  \bibinfo{journal}{Phys. Rev. Lett.} \textbf{\bibinfo{volume}{99}},
  \bibinfo{pages}{250403} (\bibinfo{year}{2007}).

\bibitem[{\citenamefont{Liao et~al.}(2010)\citenamefont{Liao, Rittner,
  Paprotta, Li, Partridge, Hulet, Baur, and Mueller}}]{Liao_10}
\bibinfo{author}{\bibfnamefont{Y.~A.} \bibnamefont{Liao}},
  \bibinfo{author}{\bibfnamefont{A.~S.~C.} \bibnamefont{Rittner}},
  \bibinfo{author}{\bibfnamefont{T.}~\bibnamefont{Paprotta}},
  \bibinfo{author}{\bibfnamefont{W.}~\bibnamefont{Li}},
  \bibinfo{author}{\bibfnamefont{G.~B.} \bibnamefont{Partridge}},
  \bibinfo{author}{\bibfnamefont{R.~G.} \bibnamefont{Hulet}},
  \bibinfo{author}{\bibfnamefont{S.~K.} \bibnamefont{Baur}}, \bibnamefont{and}
  \bibinfo{author}{\bibfnamefont{E.~J.} \bibnamefont{Mueller}},
  \bibinfo{journal}{Nature} \textbf{\bibinfo{volume}{467}},
  \bibinfo{pages}{567} (\bibinfo{year}{2010}).

\bibitem[{\citenamefont{Carlson et~al.}(2000)\citenamefont{Carlson, Orgad,
  Kivelson, and Emery}}]{Carlson_00}
\bibinfo{author}{\bibfnamefont{E.~W.} \bibnamefont{Carlson}},
  \bibinfo{author}{\bibfnamefont{D.}~\bibnamefont{Orgad}},
  \bibinfo{author}{\bibfnamefont{S.~A.} \bibnamefont{Kivelson}},
  \bibnamefont{and} \bibinfo{author}{\bibfnamefont{V.~J.} \bibnamefont{Emery}},
  \bibinfo{journal}{Phys.~Rev.~B} \textbf{\bibinfo{volume}{\textbf{62}}},
  \bibinfo{pages}{3422} (\bibinfo{year}{2000}).

\bibitem[{\citenamefont{Biermann et~al.}(2001)\citenamefont{Biermann, Georges,
  Lichtenstein, and Giamarchi}}]{Biermann_01}
\bibinfo{author}{\bibfnamefont{S.}~\bibnamefont{Biermann}},
  \bibinfo{author}{\bibfnamefont{A.}~\bibnamefont{Georges}},
  \bibinfo{author}{\bibfnamefont{A.}~\bibnamefont{Lichtenstein}},
  \bibnamefont{and}
  \bibinfo{author}{\bibfnamefont{T.}~\bibnamefont{Giamarchi}},
  \bibinfo{journal}{Phys.~Rev.~Lett.} \textbf{\bibinfo{volume}{\textbf{87}}},
  \bibinfo{pages}{276405} (\bibinfo{year}{2001}).

\bibitem[{\citenamefont{Ho et~al.}(2004)\citenamefont{Ho, Cazalilla, and
  Giamarchi}}]{Ho_04}
\bibinfo{author}{\bibfnamefont{A.~F.} \bibnamefont{Ho}},
  \bibinfo{author}{\bibfnamefont{M.~A.} \bibnamefont{Cazalilla}},
  \bibnamefont{and}
  \bibinfo{author}{\bibfnamefont{T.}~\bibnamefont{Giamarchi}},
  \bibinfo{journal}{Phys.~Rev.~Lett.} \textbf{\bibinfo{volume}{\textbf{92}}},
  \bibinfo{pages}{130405} (\bibinfo{year}{2004}).

\bibitem[{\citenamefont{Kollath et~al.}(2008)\citenamefont{Kollath, Meyer, and
  Giamarchi}}]{Kollath_08}
\bibinfo{author}{\bibfnamefont{C.}~\bibnamefont{Kollath}},
  \bibinfo{author}{\bibfnamefont{J.~S.} \bibnamefont{Meyer}}, \bibnamefont{and}
  \bibinfo{author}{\bibfnamefont{T.}~\bibnamefont{Giamarchi}},
  \bibinfo{journal}{Phys.~Rev.~Lett.} \textbf{\bibinfo{volume}{\textbf{100}}},
  \bibinfo{pages}{130403} (\bibinfo{year}{2008}).

\bibitem[{\citenamefont{Chaikin and Lubensky}(1995)}]{Chaikin}
\bibinfo{author}{\bibfnamefont{P.~M.} \bibnamefont{Chaikin}} \bibnamefont{and}
  \bibinfo{author}{\bibfnamefont{T.~C.} \bibnamefont{Lubensky}},
  \emph{\bibinfo{title}{Principles of Condensed Matter Physics}}
  (\bibinfo{publisher}{Cambridge University Press}, \bibinfo{year}{1995}).

\bibitem[{\citenamefont{Kr\"uger and Scheidl}(2002)}]{PhysRevLett.89.095701}
\bibinfo{author}{\bibfnamefont{F.}~\bibnamefont{Kr\"uger}} \bibnamefont{and}
  \bibinfo{author}{\bibfnamefont{S.}~\bibnamefont{Scheidl}},
  \bibinfo{journal}{Phys. Rev. Lett.} \textbf{\bibinfo{volume}{89}},
  \bibinfo{pages}{095701} (\bibinfo{year}{2002}).

\bibitem[{\citenamefont{Orso}(2007)}]{PhysRevLett.98.070402}
\bibinfo{author}{\bibfnamefont{G.}~\bibnamefont{Orso}}, \bibinfo{journal}{Phys.
  Rev. Lett.} \textbf{\bibinfo{volume}{98}}, \bibinfo{pages}{070402}
  (\bibinfo{year}{2007}).

\bibitem[{\citenamefont{Olshanii}(1998)}]{PhysRevLett.81.938}
\bibinfo{author}{\bibfnamefont{M.}~\bibnamefont{Olshanii}},
  \bibinfo{journal}{Phys. Rev. Lett.} \textbf{\bibinfo{volume}{81}},
  \bibinfo{pages}{938} (\bibinfo{year}{1998}).

\bibitem[{\citenamefont{Samokhin}(2010)}]{PhysRevB.81.224507}
\bibinfo{author}{\bibfnamefont{K.~V.} \bibnamefont{Samokhin}},
  \bibinfo{journal}{Phys. Rev. B} \textbf{\bibinfo{volume}{81}},
  \bibinfo{pages}{224507} (\bibinfo{year}{2010}).

\bibitem[{\citenamefont{Bardeen et~al.}(1969)\citenamefont{Bardeen, K\"ummel,
  Jacobs, and Tewordt}}]{PhysRev.187.556}
\bibinfo{author}{\bibfnamefont{J.}~\bibnamefont{Bardeen}},
  \bibinfo{author}{\bibfnamefont{R.}~\bibnamefont{K\"ummel}},
  \bibinfo{author}{\bibfnamefont{A.~E.} \bibnamefont{Jacobs}},
  \bibnamefont{and} \bibinfo{author}{\bibfnamefont{L.}~\bibnamefont{Tewordt}},
  \bibinfo{journal}{Phys. Rev.} \textbf{\bibinfo{volume}{187}},
  \bibinfo{pages}{556} (\bibinfo{year}{1969}).

\bibitem[{\citenamefont{Nagaosa}(1999)}]{Nagaosa}
\bibinfo{author}{\bibfnamefont{N.}~\bibnamefont{Nagaosa}},
  \emph{\bibinfo{title}{Quantum Field Theory in Condensed Matter Physics}}
  (\bibinfo{publisher}{Springer}, \bibinfo{year}{1999}).

\bibitem[{\citenamefont{Matsuo et~al.}(1998)\citenamefont{Matsuo, Higashitani,
  Nagato, and Nagai}}]{Matsou_98}
\bibinfo{author}{\bibfnamefont{S.}~\bibnamefont{Matsuo}},
  \bibinfo{author}{\bibfnamefont{S.}~\bibnamefont{Higashitani}},
  \bibinfo{author}{\bibfnamefont{Y.}~\bibnamefont{Nagato}}, \bibnamefont{and}
  \bibinfo{author}{\bibfnamefont{K.}~\bibnamefont{Nagai}},
  \bibinfo{journal}{J.Phys.Soc.Jpn.} \textbf{\bibinfo{volume}{67}},
  \bibinfo{pages}{280} (\bibinfo{year}{1998}).

\bibitem[{\citenamefont{Yoshida and Yip}(2007)}]{PhysRevA.75.063601}
\bibinfo{author}{\bibfnamefont{N.}~\bibnamefont{Yoshida}} \bibnamefont{and}
  \bibinfo{author}{\bibfnamefont{S.-K.} \bibnamefont{Yip}},
  \bibinfo{journal}{Phys. Rev. A} \textbf{\bibinfo{volume}{75}},
  \bibinfo{pages}{063601} (\bibinfo{year}{2007}).

\bibitem[{\citenamefont{Zhao and Liu}(2008)}]{PhysRevA.78.063605}
\bibinfo{author}{\bibfnamefont{E.}~\bibnamefont{Zhao}} \bibnamefont{and}
  \bibinfo{author}{\bibfnamefont{W.~V.} \bibnamefont{Liu}},
  \bibinfo{journal}{Phys. Rev. A} \textbf{\bibinfo{volume}{78}},
  \bibinfo{pages}{063605} (\bibinfo{year}{2008}).

\end{thebibliography}

\end{document}